\documentclass[twocolumn,showpacs,preprintnumbers,amsmath,amssymb,prl,aps]{revtex4-1}

\usepackage{epsfig}
\usepackage{graphicx}
\usepackage{bm}
\usepackage{float} 
\usepackage{amsmath}
\begin{document}

\title{A review of the electronic structure of CaFe$_2$As$_2$
and FeTe$_{0.6}$Se$_{0.4}$}

\author{Kalobaran Maiti}
\altaffiliation{Corresponding author: kbmaiti@tifr.res.in}

\affiliation{Department of Condensed Matter Physics and Materials
Science, Tata Institute of Fundamental Research, Homi Bhabha Road,
Colaba, Mumbai - 400 005, INDIA.}

\date{\today}

\begin{abstract}
Fe-based superconductors have drawn much attention during the last
decade due to the finding of superconductivity in materials
containing the magnetic element, Fe, and the coexistence of
superconductivity \& magnetism. Extensive study of the electronic
structure of these systems suggested dominant role of $d$ states in
their electronic properties, whereas the cuprate superconductors
show major role of the ligand derived states. In this article, we
review some of our results on the electronic structure of these
fascinating systems employing high resolution photoemission
spectroscopy. The combined effect of electron correlation and
covalency reveal an interesting scenario in their electronic
structure. The ligand $p$ states contribution at the Fermi level is
found to be much more significant than that indicated in earlier
studies. Temperature evolution of the energy bands reveals signature
of transition akin to Lifshitz transition in these systems.
\end{abstract}


\maketitle

\section{Introduction}

High temperature superconductivity continued to be one of the thrust
area in condensed matter research for many decades, where most of
the focus has been centered around the study of cuprates
superconductors \cite{cuprate}. Discovery of superconductivity in
Fe-based compounds \cite{Kamihara1,Kamihara2} renewed great
attention in the study of high temperature superconductivity.
Fe-based systems are significantly different from the cuprates. The
parent compounds in cuprates are antiferromagnetic Mott insulators,
where the insulating property arises due to strong electron
correlation compared to the width of their conduction band. The
antiferromagnetism gets suppressed with the charge carrier doping
and superconductivity emerges beyond some critical doping. The
normal phase of these materials exhibit plethora of unusual behavior
such as pseudogap phase, strange metallicity etc.

On the other hand, the parent compounds of Fe-based compounds
(pnictides or chalcogenides) are metals exhibiting spin density wave
(SDW) phase in the ground state. Charge carrier doping in these
systems leads to superconductivity via suppression of long range
magnetic order in the parent compositions \cite{Iron-Pnictide}.
Interestingly, many of these Fe-based compounds exhibit pressure
induced superconductivity \cite{pres1,pres2}. Application of
pressure usually renormalizes the hopping interaction strengths due
to the compression/distortion of the real lattice without
significant change in the overall carrier concentration. Thus, the
finding of pressure induced superconductivity expands the domain of
unresolved puzzles significantly. Most interestingly, some Fe-based
compounds exhibit an unusual coexistence of magnetic order and
superconductivity \cite{pres2,EuFe2As2}.

Here, we review the experimental results of two Fe-based
superconductors, Fe(TeSe) and CaFe$_2$As$_2$ belonging to two
difference class of materials published earlier
\cite{ganesh_fetese,ganesh_CaFe2As2,kbm_conf} and try to bring out
common features among these materials. Fe(TeSe) group of compounds,
popularly known as `11' systems forms in anti-PbO-type crystal
structure (space group $P4/nmm$) \cite{structure}, and are believed
to be the most correlated ones due to their large `chalcogen height'
\cite{Pnictogen height} (the height of the anions from the Fe-plane)
\cite{review}. The end members, FeTe exhibits spin density wave
(SDW)-type antiferromagnetic transition at 65 K
\cite{FeTe_dos,FeTe_sdw} and FeSe is a superconductor below 8 K
\cite{FeSe1,FeSe2}. Homovalent substitution of Te at Se-sites in
FeSe leads to an increase in superconducting transition temperature,
$T_c$ with maximum $T_c$ of 15 K for 60\% Te-substitutions
\cite{CSYadav1}, despite the fact that such substitutions often
introduces disorder in the system that is expected to reduce the
superconducting transition temperature, $T_c$ \cite{nandini}.

CaFe$_2$As$_2$ belong to another class of materials known as `122'
compounds and crystallize in the ThCr$_2$Si$_2$ type tetragonal
structure at room temperature, (space group $I4/mmm$).
CaFe$_2$As$_2$ exhibits SDW transition due to the long range
magnetic ordering of the Fe moments at $T_{SDW}$ = 170 K along with
a structural transition to an orthorhombic phase. High pressure
\cite{pres1}, substitution of Fe by Co, Ni\cite{Thamiz} and other
dopants induces superconductivity in CaFe$_2$As$_2$. The SDW
transition is found to accompany a nesting of the Fermi surface
\cite{FSNesting,dessau} along with a transition from two dimensional
(2D) to three dimensional (3D) Fermi surface associated with the
structural transition \cite{3Dto2D-Kaminskii,Fink,Fink-EPL10}.

It is believed that Fe 3$d$ states play dominant role in the
electronic properties of these systems in contrast to cuprates,
where the doped holes possess dominant ligand 2$p$ orbital character
\cite{cuprate}. Thus, the physics of unconventional superconductors
is complex due to the significant differences among different
classes of materials. Here, we show that the ligand $p$ electrons
play much more important role than what was anticipated. The
temperature evolution of the electronic structure reveals
interesting scenario in these materials.

\section{Experimental}

The single crystals of CaFe$_2$As$_2$ were grown using Sn flux and
the single crystalline sample of FeTe$_{0.6}$Se$_{0.4}$
\cite{CSYadav2} was grown by self flux method. The grown crustals
were characterized by $x$-ray diffraction, Laue, M\"{o}ussbauer and
tunneling electron microscopic measurements establishing
stoichiometric and homogeneous composition of the sample with no
trace of additional Fe in the material. The grown crystals are flat
platelet like, which can be cleaved easily and the cleaved surface
looked mirror shiny. Photoemission measurements were carried out
using a Gammadata Scienta analyzer, R4000 WAL and monochromatic
photon sources, Al $K\alpha$ ($h\nu$ = 1486.6 eV), He {\scriptsize
I} ($h\nu$ = 21.2 eV) and He {\scriptsize II} ($h\nu$ = 40.8 eV)
sources. The energy resolution and angle resolution were set to 2
meV and 0.3$^o$ respectively for ultraviolet photoemission (UP)
studies and the energy resolution was fixed to 350 meV for $x$-ray
photoemission (XP) measurements. The sample was cleaved {\it in
situ} (base pressure $<~3\times$~10$^{-11}$ Torr) at each
temperature several times to have a clean well ordered surface for
the photoemission studies. Reproducibility of the data in both
cooling and heating cycle was observed. The energy band structure of
CaFe$_2$As$_2$ and FeTe$_{0.5}$Se$_{0.5}$ was calculated using full
potential linearized augmented plane wave method within the local
density approximation (LDA) using Wien2k software \cite{wien2k}. The
energy convergence was achieved using 512 $k$-points within the
first Brillouin zone.

\section{Results and Discussions}

\begin{figure}
 \vspace{-2ex}
\begin{center}
\includegraphics[scale=0.3]{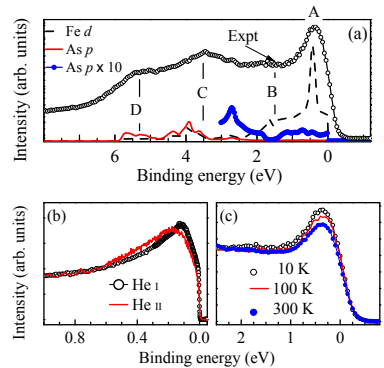}
\end{center}
\vspace{-18ex}
 \caption{(a) XP valence band spectrum of CaFe$_2$As$_2$ at 300 K (symbols).
Fe 3$d$ and As 4$p$ contributions obtained from ab initio
calculations are shown by dashed and solid lines. The solid circles
represent rescaled As 4$p$ contributions. (b) Near Fermi level
feature from He {\scriptsize I} and He {\scriptsize II} excitations.
(c) Evolution of the XP valence band spectra near $\epsilon_F$ with
temperatures.}
 \vspace{-2ex}
\end{figure}

As discussed above, the major difference of Fe-based superconductors
with the cuprates is believed to be the character of the conduction
electrons. In cuprates, the charge transfer energy (= the energy
required to transfer an electron from ligand to the copper site) is
smaller than the electron correlation strength \cite{kbm_srcuo}.
Therefore, the Fermi level, $\epsilon_F$ lies at the top of the O
2$p$ band and the electrons close to the Fermi level possess
dominant O 2$p$ character. The electron correlation strength in
Fe-pnictides is expected to be relatively smaller \cite{Silke} and
the electrons close to $\epsilon_F$ were described to be dominated
by Fe 3$d$ character. This appears to be the case in the valence
band spectrum of CaFe$_2$As$_2$ shown in Fig. 1(a). The $x$-ray
photoemission (XP) spectrum at 300 K exhibits four distinct
features, A, B, C and D \cite{ganesh_CaFe2As2}. The calculated
spectral functions obtained by convoluting the the band structure
results with the Fermi-Dirac function and resolution broadening
function exhibit good representation of the experimental spectra. It
is clear that the feature A possesses dominant Fe 3$d$ contributions
(dashed line), while the As 4$p$ states (solid line) contribute at
higher binding energies.

In order to learn this better, we critically investigate the
electronic states close to $\epsilon_F$. It appears that the
hybridization between As 4$p$ and Fe 3$d$ states is quite strong
with significant contribution coming from As 4$p$ PDOS near
$\epsilon_F$ as shown by solid circles in the figure. This is
further examined experimentally in Fig. 1(b) by comparing the
spectra obtained using He {\scriptsize I} and He {\scriptsize II}
excitations. The atomic photoemission cross-section for Fe 3$d$ and
As 4$p$ at He {\scriptsize I} energy excitation are 4.833 \& 3.856,
and at He {\scriptsize II} are 8.761 \& 0.2949, respectively
\cite{yeh}. Thus, relative intensity corresponding to As 4$p$ states
will increase significantly at He {\scriptsize I} energy compared to
He {\scriptsize II} energy. The comparison of the He {\scriptsize I}
and He {\scriptsize II} spectra having same resolution broadening
suggests that the contribution of As 4$p$ states at the Fermi level
is indeed large compared to the band structure results shown in Fig.
1(a).

The temperature evolution of the feature A in the XP valence band
spectra is shown in Fig. 1(c) after normalizing by the spectral
intensities in the energy range beyond 1 eV binding energy. The
intensity of the feature A exhibits gradual enhancement with the
decrease in temperature. Valence band spectra of a correlated system
exhibit signature of upper and lower Hubbard bands constituted by
the correlated electronic states (often called incoherent feature)
and a Kondo resonance feature called coherent feature appears at the
Fermi level representing the itinerant electrons. The decrease in
temperature leads to an increase in the coherent feature intensity
at the cost of incoherent features \cite{RMP-Gabi,ce2cosi3}. While
such increase in spectral intensity can have other origin
\cite{impurity}, we strongly feel, the enhancement of the feature A
with the decrease in temperature shown in Fig. 1(c) can be
attributed to correlation induced effects as justified later in the
text.

\begin{figure}
 \vspace{-2ex}
\begin{center}
\includegraphics[scale=0.3]{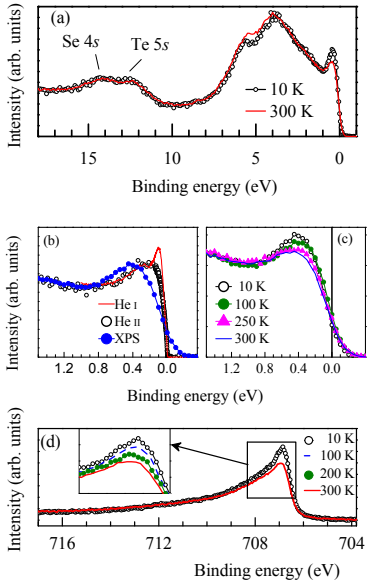}
\end{center}
 \caption{(a) XP valence band spectrum of FeTe$_{0.6}$Se$_{0.4}$ at
300 K (line) and 10 K (symbols). (b) Near Fermi level spectra at 10
K obtained using He {\scriptsize I}, He {\scriptsize II} and Al
$K\alpha$ lines at 10 K. (c) Temperature evolution of the near Fermi
level XP feature. (d) Fe 2$p$ core level spectra. Inset shows
expanded view of the well screened feature.}
 \vspace{-2ex}
\end{figure}

In Fig. 2(a), we show the valence band spectra of
FeTe$_{0.6}$Se$_{0.4}$ obtained using Al $K\alpha$ $x$-ray source at
10 K and 300 K \cite{ganesh_fetese}. The valence band exhibit
multiple features - the feature close to $\epsilon_F$ possess
dominant Fe 3$d$ character and Te/Se $p$ related features primarily
contribute in the energy range 2 - 7 eV as also appeared in the case
of CaFe$_2$As$_2$. The higher binding energy features correspond to
Se/Te $s$ states excitations. A comparison of the near Fermi level
feature at 10 K obtained using different photon energies exhibit
interesting scenario. The feature around 100 meV is most intense in
the He {\scriptsize I} spectra indicating again large chalcogen $p$
contributions. Decrease in temperature leads to a gradual increase
in intensity of the near $\epsilon_F$ feature as shown in Fig. 2(c)
consistent with the scenario in correlated electron systems.

The correlation induced effect can be verified further by inspecting
the Fe 2$p$ core level spectra shown in Fig. 2(d). The spectra
exhibit an interesting evolution with temperature. The intensity of
the peak around 707 eV binding energy increases gradually with the
decrease in temperature (see inset). This feature is often referred
as the well screened feature, where the core hole created by
photoemission is screened by a conduction electron in the final
state \cite{acker,takahashi}. Since the decrease in temperature
leads to an enhancement of the coherent feature intensity with a
consequent decrease in the incoherent feature intensity
\cite{RMP-Gabi}, the core hole is expected to be more efficiently
screened at lower temperatures as more number of mobile electrons
are available at low temperatures. Therefore, the temperature
evolution of the core level spectra observed here can also be
attributed to the correlation induced effect discussed for the
valence band.

\begin{figure}
\begin{center}
\includegraphics [scale=0.35]{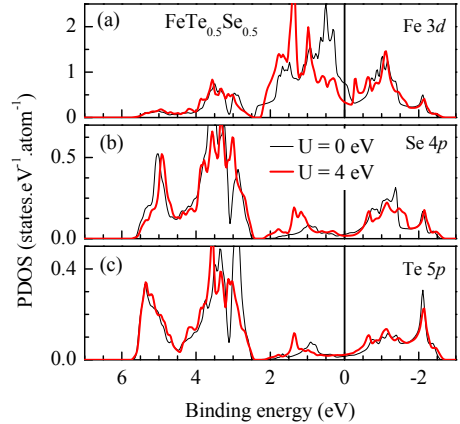}
\end{center}
 \vspace{-20ex}
\caption{Calculated (a) Fe 3$d$, (b) Se 4$p$ and (c) Te 5$p$ partial
density of states for uncorrelated (thin line for $U$ = 0 eV) and
correlated system (thick line for $U$ = 4.0 eV).}
\end{figure}

In order to investigate the dominance of $p$ character near
$\epsilon_F$ in the experimental spectra in contrast to the
prediction of the dominance of Fe 3$d$ states, we show the
calculated partial density of states corresponding to the correlated
and uncorrelated ground states \cite{ganesh_fetese}. Finite electron
correlation leads to a spectral weight transfer from $\epsilon_F$ to
higher binding energies leading to an enhancement of the intensity
around 2 eV (incoherent feature). Electron correlation affects the
electronic states with different orbital character differently
depending on their degree of itineracy in the uncorrelated system
\cite{y2ir2o7}. This is evident in Fig. 3 exhibiting significant
transfer of the Fe 3$d$ partial density of states (PDOS) to higher
binding energies. However, the Se 4$p$~/~Te 5$p$ contributions
increase near $\epsilon_F$. Thus, the relative intensity of the
$p$-states near $\epsilon_F$ will be enhanced significantly with
respect to the Fe $d$ states. This explains the presence of dominant
$p$ character near $\epsilon_F$ in the experiment.

\begin{figure}
\begin{center}
\includegraphics [scale=0.4]{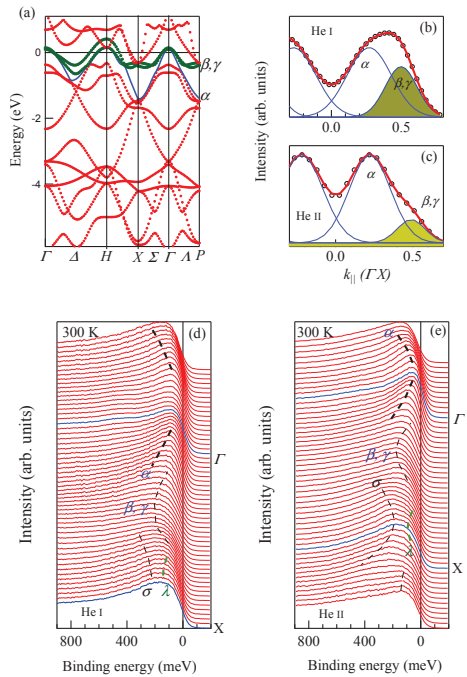}
\end{center}
 \vspace{-10ex}
\caption{(a) Calculated energy band structure of CaFe$_2$As$_2$
showing three energy bands $\alpha$, $\beta$ and $\Gamma$ making
three hole pockets around $\Gamma$ point. Momentum distribution
curves at 140 meV and 300 K in the (b) He {\scriptsize I} and (c) He
{\scriptsize II} spectra. The lines show a typical fit exhibiting
signature $\alpha$, $\beta$ and $\Gamma$ bands. Energy distribution
curves as (d) He {\scriptsize I} and (e) He {\scriptsize II} photon
energies and 300 K.}
\end{figure}

Being investigated the character of the electronic states near
$\epsilon_F$, we now turn to the energy band structure of a typical
pnictide, CaFe$_2$As$_2$ - all the samples in this class of
materials exhibit essentially similar electronic structure. The
calculated energy bands are shown in Fig. 4(a) exhibiting $t_{2g}$
bands close to the Fermi level, and both bonding \& anti-bonding
$e_g$ bands appear away from $\epsilon_F$. Three energy bands having
$t_{2g}$ symmetry and denoted by $\alpha$, $\beta$ and $\gamma$ in
the figure cross $\epsilon_F$ near $\Gamma$ point forming three
hole-pockets. Here, $\Gamma$ and $X$ points are defined as (0,0) and
($\pi$,$\pi$) in the $xy$-plane. The $k_z$ values corresponding to
He {\scriptsize I} and He {\scriptsize II} photon energies are
($k_z~\sim~$9.5$\pi/c$ and $\sim~$12.5$\pi/c$, respectively.

\begin{figure}
 \vspace{-2ex}
 \begin{center}
\includegraphics[scale=0.4]{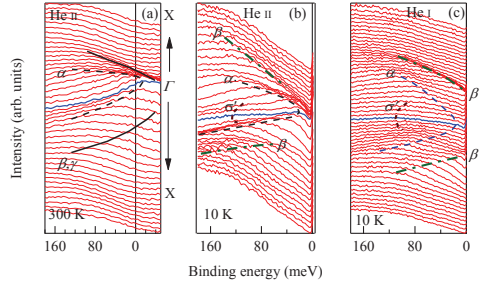}
\end{center}
 \vspace{-38ex}
 \caption{Energy distribution curves (EDCs) in the He {\scriptsize II}
at (a) 300 K and (b) 10 K. (c) The EDCs in the He {\scriptsize I}
spectra at 10 K.}
 \vspace{-2ex}
\end{figure}

Various angle resolved photoemission spectroscopic measurements
\cite{3Dto2D-Kaminskii,Fink,Fink-EPL10} show that the Fermi surface
corresponding to all these three bands exhibit two-dimensional
topology at room temperature, where the sample has tetragonal
structure. The signature of these three bands are observed in the
momentum distribution curves (MDCs) at 300 K in the He {\scriptsize
I} and He {\scriptsize II} spectra shown in Figs. 2(b) and 2(c),
respectively. The $\beta$ and $\gamma$ bands possessing
$d_{xz},d_{yz}$ symmetry appear almost degenerate, while the
$\alpha$ band having $d_{xy}$ symmetry distinctly appear at slightly
higher binding energies \cite{ganesh_CaFe2As2}. The energy
distribution curves (EDCs) in Figs. 2(d) and 2(e) show that all
these energy bands cross, $\epsilon_F$ in the vicinity of $\Gamma$
point indicating presence of three hole-pockets at room temperature.
An energy band $\lambda$ is also observed forming an electron pocket
around $X$-point. The Fermi surfaces corresponding to $\gamma$ and
$\lambda$ bands are nested leading to the SDW transition in these
materials \cite{FSNesting,dessau}.

With the decrease in temperature below the SDW transition
temperature, the $\gamma$ band hole pocket and the $\lambda$ band
electron pocket vanishes opening up a gap at the Fermi level
\cite{Fink,Fink-EPL10}. Subsequently, the crystal structure also
changes from tetragonal to orthorhombic that leads to a change in
the Fermi surface topology - the Fermi surface corresponding to the
$\alpha$ band exhibit $k_z$-dependence indicating its transition to
three dimensionality \cite{3Dto2D-Kaminskii}. It is observed that in
the orthorhombic phase, the $\alpha$-band hole pocket is centered
around $k_z~\sim~2(2n+1)\pi/c$ and it is absent around $4n\pi/c$ in
the $k$-plane containing $k_z$ axis. Thus, $\alpha$ band is expected
to cross $\epsilon_F$ at He {\scriptsize I} energy, while it will
appear below $\epsilon_F$ at He {\scriptsize II} energy. We show the
He {\scriptsize II} EDCs at 300 K and 10 K in Figs. 5(a) and 5(b),
respectively exhibiting exactly the same scenario. Interestingly,
the $\alpha$ band in the He {\scriptsize I} spectra also appears
below $\epsilon_F$ at 10 K as shown in Fig. 5(c) indicating the
vanishing of the Fermi surface corresponding to the $\alpha$ band at
10 K - larger intensity of the $\alpha$ band in He {\scriptsize II}
spectra compared to that in He {\scriptsize I} spectra indicate its
dominant Fe 3$d$ character.

It is to note here that many of the unconventional superconductors
exhibit signature of Lifshitz transition \cite{thomas}. If the Fermi
level is in proximity to a point separating hole- and electron-type
Fermi surfaces, a small change in a tuning parameter such as doping,
pressure would lead to a transition from an electron-type to
hole-type Fermi surface or vise versa. This is known as Lifshitz
transition. Proximity to Lifshitz transition indicates significant
quantum fluctuation in the system. The signature of Lifshitz
transition has been observed due to subtle change in charge carrier
concentration in cuprates \cite{thomas} as well as in electron doped
Ba(Fe$_{1-x}$Co$_x$)$_2$As$_2$ \cite{Lifschitz}. The vanishing of
the hole Fermi surface corresponding to the $\alpha$ in
CaFe$_2$As$_2$ as a function of temperature is interesting and akin
to the scenario in Lifshitz transition. These results indicate
importance of Lifshitz transition in such unconventional
superconductors.

\section{Conclusions}

In summary, we presented here a review of our studies of the
electronic structure of Fe-based superconductors,
FeTe$_{0.6}$Se$_{0.4}$ and CaFe$_2$As$_2$. A critical analysis of
the experimental and band structure results indicates that the
electronic states close to the Fermi level deriving the electronic
properties of these materials possess significantly large $p$
character. Thus, the difference between Fe-based and cuprate
superconductors appears to be much less than what was thought
earlier. The temperature evolution of the experimental spectra
exhibit signature the enhancement of the coherent feature (Kondo
resonance feature) at lower temperatures. The angle resolved
photoemission data from CaFe$_2$As$_2$ exhibit signature of Lifshitz
transition as a function of temperature.

\end{document}